\documentclass[print,prc,showpacs,printnumbers,superscriptaddress,floatfix]{revtex4}
\usepackage{dcolumn}
\usepackage{bm}
\usepackage{longtable}
\usepackage{graphicx,epsfig,latexsym,amssymb}
\usepackage{multirow,amsmath,array,booktabs,color}
\usepackage[section]{placeins}

\begin{document}

\title{Resonant states and pseudospin symmetry in the Dirac Morse potential}
\author{Quan Liu}
\author{Zhong-Ming Niu}
\author{Jian-You Guo}
\email{jianyou@ahu.edu.cn}
\affiliation{School of Physics and Material Science, Anhui University, Hefei 230039, P.R.
China}
\date{\today }

\begin{abstract}
The complex scaling method is applied to study the resonances of a Dirac
particle in a Morse potential. The applicability of the method is
demonstrated with the results compared with the available data. It is shown
that the present calculations in the nonrelativistic limit are in excellent
agreement with the nonrelativistic calculations. Further, the dependence of
the resonant parameters on the shape of the potential is checked with the
sensitivity to the potential parameters analyzed. By comparing the energies
and widths of the pseudospin doublets, well pseudospin symmetry is
discovered in the present model. The relationship between the pseudospin
symmetry and the shape of the potential is investigated by changing the
Morse potential shaped by the dissociation energy, the equilibrium
intermolecular distance, and the positive number controlling the decay
length of the potential.
\end{abstract}

\maketitle

\section{Introduction}

It is well known that the resonances are the most striking phenomenon in the
whole range of scattering experiments, and appear widely in atomic,
molecular, nuclear physics and in chemical reactions. One of the best-known
examples is the spontaneous positron production in uranium nuclei
collisions, where some harp resonance peaks have been observed~\cite%
{Cowan1985prl}.

To explore the resonances, ones have developed many techniques, which
include the R-matrix theory~\cite{Hale1987prl}, K-matrix theory~\cite%
{Humblet1991prc}, the scattering phase shift method~\cite{Taylor1972}, and
several bound-state-like methods, such as the real stabilization method
(RSM)~\cite{Hazi1970pra}, the analytic continuation in the coupling constant
(ACCC) method~\cite{Kukulin1989} and the complex scaling method (CSM)~\cite%
{HoPR1983, MoiseyevPR1998}. Due to the simplicity in the calculations, these
bound-state-like methods have been widely employed to study resonances in
quantum many-body systems. For examples, much effort has been made in order
to calculate more efficiently resonance parameters with the RSM~\cite%
{Taylor1976, Mandelshtam1994pra, Kruppa1999pra,Zhang2008prc}. Combined with
the cluster model, the ACCC approach has been used to calculate the
resonances in some light nuclei~\cite{TanakaPRC1997, TanakaPRC1999}. In
combination with the relativistic mean field (RMF) theory~\cite%
{YangCPL2001,ZhangPRC2004,ZhangPRC2012}, the RMF-ACCC has presented a good
description for the structure of resonant levels in some realistic nuclei~%
\cite{GuoPRC2005, GuoPRC2006}. Moreover, the CSM have been successfully used
to study resonances in atomic and molecular systems~\cite{AlhaidariPRA2007,
BylickiPRA2008} and atomic nuclei~\cite{KruppaPRL1997, AraiPRC2006,
GuoPRC2010,LiuPRC2012, MyoPRC2012}.

Recently, the Morse potential~\cite{MorsePR1929} attracts additional
attentions for its success in describing the motion of atomic and molecular
as well as nuclear systems. The vibration of a diatomic molecule was well
depicted by the Morse potential~\cite{GogginPRA1988}. The Morse potential is
also very useful in expressing the properties of nuclei~\cite%
{BoztosunPRC2008,AydogduPLB2011,InciPRC2011}. Other applications of Morse
potential include the description of the interactions between two atoms in a
diatomic molecule~\cite{MoralesCPL1989, FilhoPLA2000, KillingbeckJCP2002,
DongIJQP2002}, the interatomic potential of crystalline solids~\cite%
{MohammedPRB1984}, and the potential for the adsorption of a molecule or
atom by a solid surface~\cite{SunPRB2011}. This list is far from
comprehensive and new applications continue to be found as well.

Due to the wide applicability, it is interesting to probe the resonances in
a Morse potential. Several methods have been developed to study the resonant
states of a nonrelativistic particle in a Morse potential. Nasser \emph{et al%
}.~\cite{NasserJPB2008} have adopted the J-matrix method to examine not only
the bound states but also resonances associated with the rotating Morse
potential model. Satpathy \emph{et al}.~\cite{SatpathyJPG1990,
SatpathyPPNP1992, SatpathyJPG1994, SatpathyJPG2005} have employed the
long-ranged Morse potential between two carbon nuclei in order to obtain
many vibrational states which can be compared with observed resonance
levels. Kato and Abe~\cite{KatoPRC1997} have applied the Morse potential
model to the $^{12}$C-$^{12}$C system as a typical example of the molecular
resonances observed in many lighter heavy-ion collisions. Jarukanont \emph{%
et al}.~\cite{JarukanontPRA2007} have applied the Floquet theory in
combination with the exterior complex scaling to obtain the energies and the
distributions of probability for the quasibound states of a driven Morse
potential.

In addition to these nonrelativistic studies, much attention has been paid
to the problem of the relativistic Morse potential. In 2001, the $s$-wave
Dirac-Morse problem was formulated and solved exactly~\cite{AlhaidariPRL2001}%
. As there are no analytical solutions to the Dirac equation with $\kappa
\neq 0$ for the rotational Morse potential, some approximations are employed
to present the numerical or quasi-analytical solutions~\cite{Flugge1974,
Cooper2001}. To recognize the relativistic effects of the Morse potential,
Dirac equation has been solved for attractive scalar and repulsive vector
Morse potentials under pseudospin symmetry by the Pekeris approximation~\cite%
{BerkdemirNPA2006,QiangJPA2007}. Other interesting work includes the
solution of the relativistic Morse potential problem under the condition of
pseudospin symmetry~\cite{Sever2011PLB}.

These relativistic studies on Morse potential associated with an interesting
concept, pseudospin symmetry (PSS). The concept was first introduced in the
field of nuclear physics more than 40 years ago~\cite{Hecht1969NPA,
Arima1969PLB}. Its meaning is that the single particle states with quantum
numbers $(n,l,j=l+1/2)$ and $(n-1,l+2,j=l+3/2)$ are nearly degenerate. Until
1977, this symmetry was known as a relativistic symmetry and the condition
of exact PSS was proposed in Ref.~\cite{Ginocchio1997PRL}. Afterwards, the
condition of exact PSS was extended to a more general case~\cite{Meng1998PRC}%
, and found to be approximately satisfied in exotic nuclei with highly
diffuse potentials~\cite{Meng1999PRC}. The dynamical character of the
symmetry was recognized in Ref.~\cite{Alber01} and the spin symmetry in the
antinucleon spectrum was indicated in Ref.~\cite{Zhou03}. Several recent
progresses include the supersymmetric description for deformed nuclei~%
\cite{Leviatan09}, the nonperturbation nature of PSS~\cite%
{Liang2011PRC,Gonoc11}, the relativistic effect of PSS~\cite{Chen2012PRC},
the influence of tensor interaction on PSS~\cite{Long2010PRC,Castro2012PRC},
the role of central Coulomb potential at PSS~\cite{Castro2012PRA}, and the spin and
pseudospin symmetries of the Dirac equation with confining central
potentials~\cite{Alberto2013PRC}. In Ref.~\cite{Guo2012PRCR}, the similarity
renormalization group is used as the critical tool for understanding the
origin of the PSS and the cause why the PSS becomes better for the levels
closer to the continuum is disclosed in a quantitative way~\cite{Lidp2013PRC}%
. In combination with the supersymmetric quantum mechanics, the origin of
the PSS and its breaking mechanism are detailedly analyzed at the
nonrelativistic limit~\cite{Liang2013PRC}. This symmetry is also checked in
the resonant states with similar features to
bound states indicated in Ref.~\cite{Lu2012PRL,Lu2013arXiv}, where it was
indicated that in conditions of pseudospin symmetry the same pseudospin
quantum numbers will be conserved and the pseudospin doublets would have the
same energy and width. As the interesting phenomenon also appears in the
resonant states, in this work we will apply the CSM to study the resonance
states of a Dirac particle populating in the Morse potential, and survey the
existence of PSS for this system.

\section{Theory}

The Dirac equation of describing a particle moving in the Morse potential $%
V(r)$ is
\begin{equation}
H(r)\psi (r)=\left( {\begin{array}{*{20}c} {V + M} & { - \frac{d}{{dr}} +
\frac{{ - 1 + \kappa }}{r}} \\ {\frac{d}{{dr}} + \frac{{1 + \kappa }}{r}} &
{V - M} \\ \end{array}}\right) \left( {\begin{array}{*{20}c} {f(r)} \\
{g(r)} \\ \end{array}}\right) =\varepsilon \left( {\begin{array}{*{20}c}
{f(r)} \\ {g(r)} \\ \end{array}}\right)  \label{Diraceq}
\end{equation}%
with
\begin{equation}
V(r)=V_{0}e^{-\left( r-r_{0}\right) \alpha }\left( 2-e^{-\left(
r-r_{0}\right) \alpha }\right) ,  \label{potential}
\end{equation}%
where $\kappa =\pm( j + 1/2)$ for $j =l \mp 1/2$, $V_{0}$ is the strength of
the potential (or the dissociation energy in the context of diatomic
molecules), $r_{0}$ is the equilibrium intermolecular distance, and $\alpha $
is the positive number controlling the decay length of the potential. For $%
V_{0}<0$, this potential has a minimum of $V_{0}$ at $r=r_{0}$ and it is
called the regular Morse potential. For $V_{0}>0$ , it has a maximum of $%
V_{0}$ there and it is called the inverted Morse potential, and then $r_{0}$
is the location of the top of the barrier.

The starting point of the CSM is a transformation of the Hamiltonian $H$,
which is operated by defining an unbounded nonunitary scaling operator $%
U(\theta )$ with a real parameter $\theta $. Similar to Ref.~\cite%
{GuoPRC2010,Guo2010CPC}, the complex scaling operator takes the form
\begin{equation}
U\left( \theta \right) =\left( {\begin{array}{*{20}c} {e^{i\theta \hat S} }
& 0 \\ 0 & {e^{i\theta \hat S} } \\ \end{array}}\right) ,
\end{equation}%
where $\hat{S}=r\frac{\partial }{{\partial r}}$. Then, the Dirac Hamiltonian
in Eq.~(\ref{Diraceq}) becomes a complex scaling form:
\begin{equation}
H_{\theta }=U\left( \theta \right) HU\left( \theta \right) ^{-1}=\left( {%
\begin{array}{*{20}c} {V\left( {re^{i\theta } } \right) + M} & {e^{ -
i\theta } \left( { - \frac{d}{{dr}} + \frac{{ - 1 + \kappa }}{r}} \right)}
\\ {e^{-i\theta } \left( {\frac{d}{{dr}} + \frac{{1 + \kappa }}{r}} \right)}
& {V\left( {re^{i\theta } } \right) - M} \\ \end{array}}\right) .
\end{equation}%
The corresponding complex scaled equation is
\begin{equation}
H_{\theta }\psi _{\theta }(r)=\varepsilon _{\theta }\psi _{\theta }\left(
r\right) ,  \label{CDiraceq}
\end{equation}%
where $\psi _{\theta }=U\left( \theta \right) \psi \left( {r}\right) $ is
the complex scaled wave function. According to the Aguilar-Balslev-Combes
(ABC) theorem \cite{Aguilar1971}, the energy for the bound states is
independent of $\theta $, and the real part $E_{r}$ and the imaginary $E_{i}$
part of the energy $\varepsilon _{\theta }$ for the resonant states are
independently of $\theta $ as well. ${\Gamma =-2}E_{i}$ represents the width
for the resonant states.

In this work, Eq.(\ref{CDiraceq}) is solved by expanding the large and small
components of the Dirac spinors $f(r)$ and $g(r)$ in terms of the radial
functions $R_{nl}\left( r\right) $ of a spherical harmonic oscillator (HO)
potential, i.e.,
\begin{equation}
f_{\theta }\left( r\right) =\sum\limits_{n=1}^{n_{\max }}{f_{n}\left( \theta
\right) }R_{n\,l}\left( r\right) \;;\quad g_{\theta }\left( r\right)
=\sum\limits_{n=1}^{\tilde{n}_{\max }}{g_{\tilde{n}}\left( \theta \right) }%
R_{\tilde{n}\,\tilde{l}}\left( r\right) ,  \label{Eq:fgHO}
\end{equation}%
where the orbital angular momenta $l$ and $\tilde{l}$ have the same meaning
as that in Ref.~\cite{GambhirAP1990}. The upper limits $n_{\max }$ and $%
\tilde{n}_{\max } $ in Eq.(6) are radial quantum numbers determined by the
corresponding major shell quantum numbers $N_{\max }$ $=2\left( n_{\max
}-1\right) +l$ and $\tilde{N}_{\max }$ $=2\left( \tilde{n}_{\max }-1\right) +%
\tilde{l}$. In the present calculations, the maximum of the major shell quantum number $N= 200$ is chosen to ensure the calculated result with the accuracy better than the order of $10^{-4}$ ${\rm fm}^{-2}$ or $10^{-4}$ a.u. for the energies and widths. The details can be referred to the literatures~\cite{GambhirAP1990,Guo2010CPC}.

Inserting Eq.(\ref{Eq:fgHO}) into Eq.~(\ref{CDiraceq}) and using the
orthogonality of wave functions $R_{nl}$, one arrives at a symmetric matrix
diagonalization problem of the dimension $n_{\max }+\tilde{n}_{\max }$,
i.e.,
\begin{equation}
\left( {\begin{array}{*{20}c} {V_{n,n'} + M \cdot \delta _{n,n'} } &
{B_{n,\tilde n'} } \\ {B_{\tilde n,n'} } & {V_{\tilde n,\tilde n'} - M \cdot
\delta _{\tilde n,\tilde n'} } \\ \end{array}}\right) \left( {%
\begin{array}{*{20}c} {f_{n'} } \\ {g_{\tilde n'} } \\ \end{array}}\right)
\;=\varepsilon _{\theta }\left( {\begin{array}{*{20}c} {f_n } \\ {g_{\tilde
n} } \\ \end{array}}\right) \;.
\end{equation}%
The matrix elements $V_{n,n^{\prime }}$ and $B_{n,\tilde{n}^{\prime }}$ are
given by
\begin{equation}
V_{n,n^{\prime }}=\int_{0}^{\infty }{r^{2}dr\;R_{nl}(r)\left[ {V(re^{i\theta
})}\right] R_{n^{\prime }l}(r)},
\end{equation}%
\begin{equation}
B_{\tilde{n},n^{\prime }}=e^{-i\theta }\int_{0}^{\infty }{r^{2}dr\left[ {R_{%
\tilde{n}\,\tilde{l}}\left( r\right) \left( {\frac{d}{{dr}}+\frac{{1+\kappa }%
}{r}}\right) R_{n^{\prime }\,l}\left( r\right) }\right] }.
\end{equation}
The integral in Eq.(8) is calculated with the Gauss quadrature
approximation~\cite{AlhaidariPRA2007,Guo2010CPC}. The matrix elements $B_{n,\tilde{n}%
^{\prime }}$ can be further simplified as
\begin{equation}
B_{\tilde{n},n^{\prime }}=\left\{
\begin{array}{l}
-\frac{1}{{b_{0}e^{i\theta }}}\left( {\sqrt{\tilde{n}+l+1/2}\delta _{\tilde{n%
},n^{\prime }}+\sqrt{\tilde{n}}\delta _{\tilde{n},n^{\prime }-1}}\right)
,\quad \kappa <0 \\
\frac{1}{{b_{0}e^{i\theta }}}\left( {\sqrt{\tilde{n}+l-1/2}\delta _{\tilde{n}%
,n^{\prime }}+\sqrt{\tilde{n}-1}\delta _{\tilde{n},n^{\prime }+1}}\right)
,\quad \kappa >0 \\
\end{array}%
\right.
\end{equation}%
With the matrix elements $V_{\tilde{n}\tilde{n}^{\prime }}$ and $B_{n,\tilde{%
n}^{\prime }}$, the solutions of the Dirac equation are obtained by
diagonalizing the matrix
\begin{equation}
H_{\theta }=\left( {\begin{array}{*{20}c} {V_{n,n'} + M \cdot \delta _{n,n'}
} & {B_{n,\tilde n'} } \\ {B_{\tilde n,n'} } & {V_{\tilde n,\tilde n'} - M
\cdot \delta _{\tilde n,\tilde n'} } \\ \end{array}}\right) .
\end{equation}%
The eigenvalues of $H_{\theta }$ representing bound or resonant states do
not change with $\theta $, while eigenvalues representing continuous
spectrum rotate.

\section{Result and discussion}

With the theoretical formalism represented in the previous section, we
explore the resonant states for a particle populating in a Morse potential.
For comparison with the results from the J-matrix approach in Ref.~\cite%
{NasserJPB2008}, the same values for the parameters of the Morse potential
are employed. The illustrated results are plotted in Fig.1 with the complex
rotation angle $\theta =70^{\circ }$ for the states with $\kappa =-1$. It should be mentioned that the energy displayed in Fig.1 is multiplied by a factor $2M/\hbar^{2}$ in order to compare with Ref.~\cite%
{NasserJPB2008}. The same as Fig.1, the energy and the depth of potential in the following Table I and Figs.2 and 3 are multiplied by a factor $2M/\hbar^{2}$. All the eigenvalues of $%
H_{\theta }$, which correspond to the bound, resonant, and continuum states,
are respectively labeled as open boxes, open circles, and solid circles.
From Fig.1, one sees clearly that the eigenvalues of $H_{\theta }$ fall into
three regions: the bound states populate on the negative energy axis, while
the continuous spectrum of $H_{\theta }$ rotates clockwise with the angle $%
2\theta $, and resonances in the lower half of the complex energy plane
located in the sector bound by the new rotated cut line and the positive
energy axis get exposed and become isolated. In the present calculations, as
the finite basis is used, the continuous spectrum of $H_{\theta }$ consists
of a string of points.

The energies for the resonant states in Fig.1 are listed in Table~\ref{tab1}. The relativistic results are shown in the first column. The second column
displays the data in the nonrelativistic limit, which are obtained with an
approximation by increasing the value of the speed of light up to 100 times
larger than its actual value in the relativistic CSM calculations~\cite{AlhaidariPRA2007}. In order to compare with
the other methods, some data from the nonrelativistic calculations are
exhibited in the third column by J-matrix method~\cite{NasserJPB2008} and
the fourth column by S-matrix method~\cite{Rawitscher2002}. From Table~\ref%
{tab1}, it can be seen that the deviation between the relativistic result
and that in the nonrelativistic limit is observable, which implies the
relativistic effect can not be completely ignored in the present model. The
present calculations in the nonrelativistic limit agree with the
nonrelativistic calculations quite well. The deviation between our results
and the data by J-matrix or S-matrix method is very small. Whether for a
relativistic or nonrelativistic system, the present method has provided a
good description for the resonant states. Hence, we can look into all the
resonant states for a Dirac particle in a Morse potential by this method.

To survey the resonances in a wider range, it is necessary to analyze the
dependence of the resonant parameters on the shape of the potential, which
can help us to recognize the resonances for more realistic systems. The
variations of energy and width with the parameters of the potential are
shown in Fig.2, where the data associated with the dissociation energy, the
equilibrium intermolecular distance, and the positive number controlling the
decay length of the potential are respectively shown in the panels (a), (b),
(c) of Fig.2. Starting from $V_{0}=1.5$ fm$^{-2}$, the energy of the
resonant states increases with the increasing of $V_{0}$. Further increasing
$V_{0}$, the energy level changes the direction of evolution for the $%
1s_{1/2}$, followed by the $2s_{1/2}$, and so on. Simultaneously, some new
resonant levels are exposed. For the width, its evolution with $V_{0}$ is
consistent with that for the energy. Starting from $V_{0}=1.5$ fm$^{-2}$,
the width reduces with the increasing $V_{0}$ for the several lower levels.
For the several upper levels, the width increases with the increasing of $%
V_{0}$. As $n$ increases, the width is sequentially changed in the direction
of the evolution with $V_{0}$. The phenomenon can be explained by the
potential energy curves plotted in Fig.3. With the increasing of $V_{0}$,
the barrier becoming higher and the well becoming deeper are seen in
Fig.3(a) for the Morse potential. It is the cause that the resonant
parameters vary with $V_{0}$ in accordance with the foregoing. With the
increasing of $n$, the evolution of the resonant energy with $r_{0}$ changes
from a decreasing to an increasing with the increasing $r_{0}$. The level
located on the middle, the energy of the state ($4s_{1/2}$) only has a
slight change. The energy for the levels lower than $4s_{1/2}$ decreases
with the increasing $r_{0}$, while the energy increases with the increasing $%
r_{0}$ for the energy level populating in higher than $4s_{1/2}$, which is
associated with the variation of the potential. From the potential energy
curves in Fig.3(b), one can see that both the barrier and the well become
wider with the increasing of $r_{0}$, which leads a weakening of penetration
for a particle in the Morse potential. Thus, the width becomes systemically
smaller with the increasing $r_{0}$. Fig.2(c) shows the resonant parameters
varying with $\alpha $, a similar change is seen for the energy with $r_{0}$
in Fig.2(b) and for the width with $V_{0}$ in Fig.2(a). These can be
explained from the changes in the potential field. The potential energy
curves in Fig.3(c) display that the potential well becomes deeper and the
range of potential barrier becomes narrower with the increase of $\alpha $,
which is the cause of the resonant parameters with $\alpha $.

To further examine the present CSM calculations, the resonant states with $%
\kappa =2$ is investigated with the data exhibited in Table~\ref{tab2}. For
comparisons with the data in Ref.~\cite{NasserJPB2008}, the atomic units $%
\hbar =m=1$ is adopted here. Similar to Table~\ref{tab1}, the result in the
nonrelativistic limit is obtained according to the method in Ref.~\cite%
{AlhaidariPRA2007}. From Table~\ref{tab2}, it can be seen our results are in
excellent agreement with the data in Ref.~\cite{NasserJPB2008}, which are
obtained by J-matrix method for a nonrelativistic system. These show the
present calculations is fully correct, and we can explore the resonant
states of a relativistic particle by this method.

With the resonance states obtained by the CSM, the position of pseudospin
doublets in the complex energy surface is shown in Fig.4, where the
pseudospin doublets $\kappa =-1$ and $2$, $\kappa =-2$ and $3$, $\kappa =-3$
and $4$, as well as $\kappa =-4$ and $5$ are respectively displayed in
Fig.4(a), (b), (c), and (d). It is clear that PSS is well preserved for all
the doublets. For the pseudospin doublet with the same orbital angular
momentum, the quality of PSS improves with the increasing of the radial
quantum number.

To disclose the relationship of the PSS in the resonant states and the shape
of the potential, we analyze the variation of the pseudospin splittings with
the parameters in the Morse potential. Keeping $r_{0}$ and $\alpha $ fixed,
we vary $V_{0}$ to see how the energy and width of the pseudospin doublets
are sensitive to the dissociation energy. This dependence is shown in
Fig.5(a) for the energy and width splitting. It is clear that the energy
splitting of the pseudospin doublets increase with the increasing of $V_{0}$%
, while a inverse trend is found for the variation of width. This can be
easily understood as higher barrier and deeper potential well are
accompanied with the increase of $V_{0}$. Similarly, the energy and width
splittings varying with $r_{0}$ is displayed in Fig.5(b) with the other
parameters fixed. From the left $(r_{0}=0.5)$, as the radius $r_{0}$
increases, the energy and width splittings decrease, which can be understood
by the dependence of $dV(r)/dr$ on $r$. The variation of the pseudospin
splittings with $\alpha $ is presented in Fig.5(c). For all the pseudospin
doublets, the tendency for the change of the pseudospin energy and width
splittings with $\alpha $ is the same. With $\alpha $ increases, the
splitting increases. This could be expected because the potential well
becomes deeper and widened with the increasing of $\alpha $, hence the
increase of the derivative of $V(r)$. In the range of the potential
parameters considered here, the deviations between pseudospin doublets is
within $1.17$ a.u. for $E_{r}$ and $1.04$ a.u. for $\Gamma $, which
indicates the PSS is well preserved.

\section{Summary}

In summary, the resonant states of Morse potential are investigated by using
the complex scaling method in the relativistic framework. The calculated
results are compared with the available data in references and satisfactory
agreements are found. Further, the dependence of the resonant parameters on
the shape of the potential is checked and a unified result is found for the
evolution of the resonant parameters with the potential parameters, which
can be well explained from the change of potential energy curves. With these
resonant states obtained, the PSS in the resonant states is investigated and
well PSS is found in the present model. The pseudospin splitting is shown in
correlation with the Morse potential shaped by the dissociation energy $V_{0}
$, the equilibrium intermolecular distance $r_{0}$, and the positive number
controlling the decay length of the potential $\alpha$, in which three Morse
potential parameters $V_{0}$, $\alpha $ and $r_{0}$ are found to play the
important roles in the splittings of energy and width of pseudospin doublets.

\begin{acknowledgments}
This work was partly supported by the National Natural Science Foundation of
China under Grants No. 10675001, No. 11175001, and No. 11205004; the Program
for New Century Excellent Talents in University of China under Grant No.
NCET-05-0558; the Excellent Talents Cultivation Foundation of Anhui Province
under Grant No. 2007Z018; the Natural Science Foundation of Anhui Province
under Grant No. 11040606M07; the Education Committee Foundation of Anhui
Province under Grant No. KJ2009A129; the Talent Foundation of High Education
of Anhui Province for Outstanding Youth (2011SQRL014) and the 211 Project of
Anhui University.
\end{acknowledgments}

\begin{table*}[tbp]
\caption{Energies in the present calculations for the states with $\protect%
\kappa = -1$ in comparison with the data in Refs.~\protect\cite%
{NasserJPB2008, Rawitscher2002}. Here the parameters of Morse potential are
adopted as $V_0 = 6~\text{fm}^{-2}$, $r_0 = 4~\text{fm}$, and $\protect%
\alpha = 0.3~\text{fm}^{-1}$. The mass of particle is adopted as $M=0.5~%
\text{fm}^{-1}$. For comparison with Refs.~\protect\cite{NasserJPB2008,
Rawitscher2002}), the depth of potential displayed and the calculated
energies are multiplied by a factor $2M/\hbar^{2}$. }
\label{tab1}%
\begin{tabular}{cccc}
\hline\hline
$E_r+iE_i (~\text{fm}^{-2})$ & $E_r+iE_i (~\text{fm}^{-2})$ & $E_r+iE_i (~%
\text{fm}^{-2})$ & $E_r+iE_i (~\text{fm}^{-2})$ \\ \hline
Relativistic & Nonrelativistic limit & Nasser ~\cite{NasserJPB2008} &
Rawitscher ~\cite{Rawitscher2002} \\ \hline
-8.1096 & -8.1089 & -8.1090 & -8.1090 \\
1.1745 -i 0.0002 & 1.1779 -i 0.0011 & 1.1778 -i 2.01E-13 & 1.1783 \\
5.6229 -i 0.0349 & 5.6212 -i 0.0303 & 5.6252 -i 0.0351 & 5.6252 -i 0.0351 \\
6.8906 -i 1.3170 & 6.9041 -i 1.3103 & 6.8911 -i 1.3194 &  \\
7.3194 -i 3.5858 & 7.3375 -i 3.6189 & 7.3182 -i 3.5887 &  \\
7.1143 -i 6.0693 & 7.0545 -i 6.0853 & 7.1111 -i 6.0715 &  \\
6.3679 -i 8.5999 & 6.3820 -i 8.5374 & 6.3627 -i 8.6005 &  \\
5.1514 -i 11.0980 & 5.1806 -i 11.1429 & 5.1446 -i 11.0960 &  \\
3.5200 -i 13.5209 & 3.4649 -i 13.5147 & 3.5123 -i 13.5151 &  \\
1.5173 -i 15.8439 & 1.5262 -i 15.8071 & 1.5095 -i 15.8334 &  \\
-0.8212 -i 18.0518 & -0.8167 -i 18.0490 & -0.8278 -i 18.0358 &  \\
-3.4656 -i 20.1350 & -3.4773 -i 20.1177 & -3.4697 -i 20.1128 &  \\
-6.3907 -i 22.0869 & -6.3930 -i 22.0537 & -6.3907 -i 22.0579 &  \\
-9.5750 -i 23.9031 & -9.5671 -i 23.8658 & -9.5691 -i 23.8667 &  \\
-12.9998 -i 25.5801 & -12.9859 -i 25.5370 & -12.9861 -i 25.5363 &  \\
-16.6491 -i 27.1155 & -16.6258 -i 27.0640 & -16.6255 -i 27.0642 &  \\
-20.5088 -i 28.5077 & -20.4731 -i 28.4487 &  &  \\ \hline\hline
\end{tabular}%
\end{table*}

\begin{table}[tbp]
\caption{The same as TABLE I, but for the states with $\protect\kappa=2$ in
comparison with the data in Ref.\protect\cite{NasserJPB2008}. Here the
parameters of Morse potential are adopted as $V_0 = 10$, $r_0 = 1$, and $%
\protect\alpha = 2.0$ in the atomic units $\hbar=m=1$.}
\label{tab2}%
\begin{ruledtabular}
\begin{tabular}{ccc}
$E_r+iE_i$                         &$E_r+iE_i$                         &$E_r+iE_i$ \\
\hline
Relativistic                       &Nonrelativistic limit              &Nasser ~\cite{NasserJPB2008}\\
\hline
-30.7047                     &-30.4136                     &-30.4139                \\
 10.8020 -i 0.2822        & 10.9262 -i 0.3026        &  10.9260 -i 0.3027    \\
 17.1419 -i 12.3689        & 17.1244 -i 12.5031        &  17.1240 -i 12.5027   \\
 11.1795 -i 32.0868        & 11.0511 -i 32.1914        &  11.0521 -i 32.1906   \\
 -4.8377 -i 52.5443         & -5.0383 -i 52.5395        &  -5.0376-i 52.5407   \\
-29.3283 -i 72.2381        &-29.5208 -i 72.0565        &                           \\
\end{tabular}
\end{ruledtabular}
\end{table}

\begin{figure}[tbp]
\includegraphics[width=10cm]{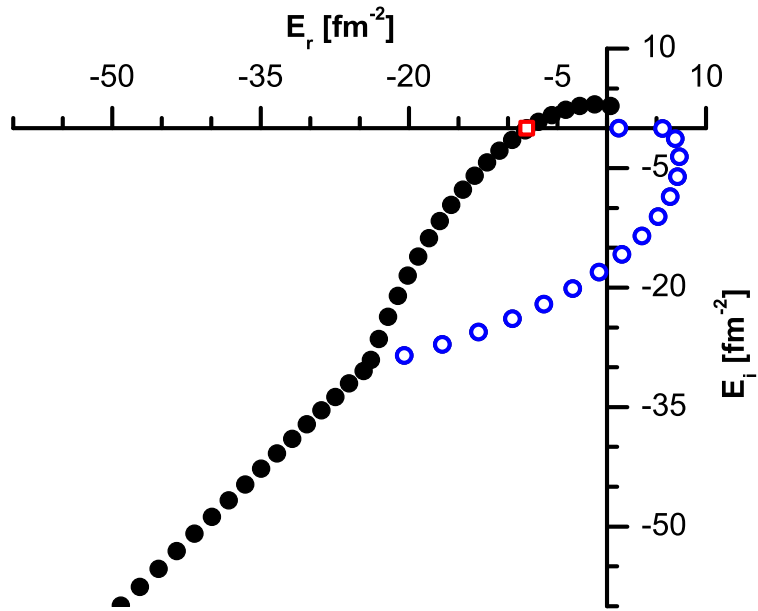}
\caption{(Color online) Position in complex energy surface for the states
with $\protect\kappa=-1$, where the bound, the resonant, and the continuum
are respectively labeled as open boxes, open circles, and solid circles. The
parameters in the Morse potential are adopted as $V_0 = 6~\text{fm}^{-2}$, $%
r_0 = 4~\text{fm}$, and $\protect\alpha = 0.3~\text{fm}^{-1}$. The mass of
particle $M=0.5~\text{fm}^{-1}$ and the rotation angle $\protect\theta =
70^\circ$ are adopted in the present calculations.}
\label{fig1}
\end{figure}

\begin{figure}[tbp]
\includegraphics[width=14cm]{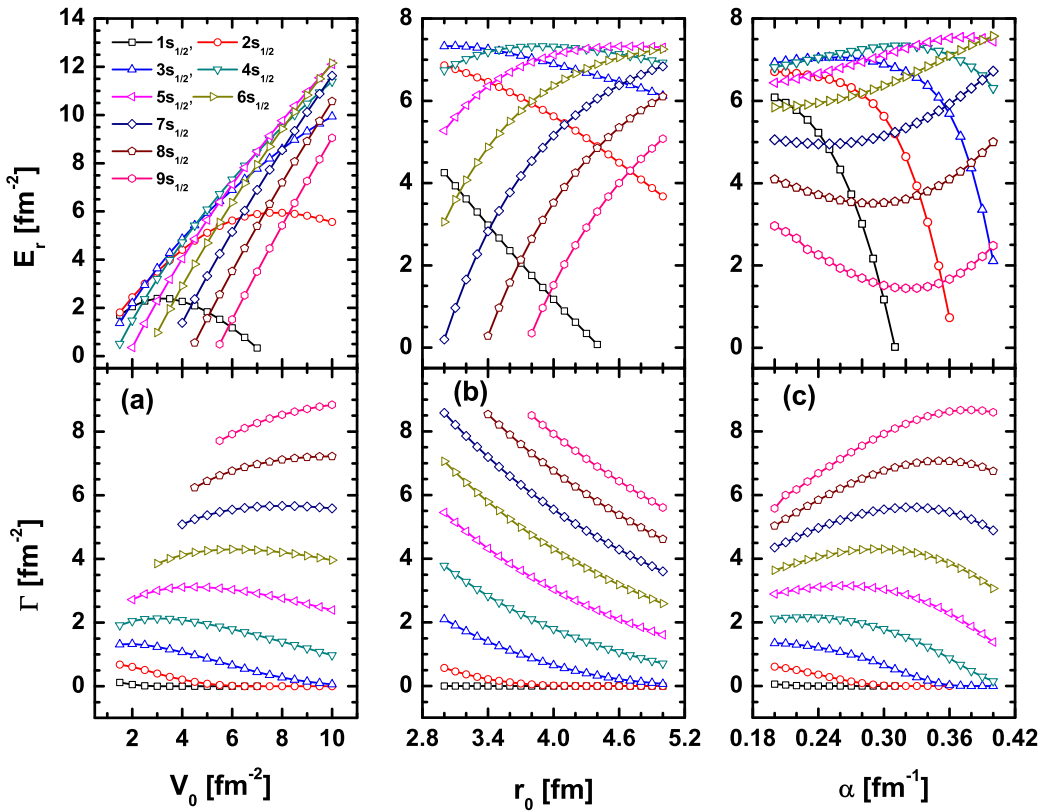}
\caption{(Color online) Energies and widths as a function of every potential
parameter for the resonant states shown in Fig.~1. The data corresponding to
the variables $V_0$, $r_0$, and $\protect\alpha$ are respectively displayed
in the subfigures (a), (b), and (c) with the other parameters fixed to $r_0
= 4~\text{fm}$, $\protect\alpha = 0.3~\text{fm}^{-1}$ in (a), $V_0 = 6~\text{%
fm}^{-2}$, $\protect\alpha = 0.3~\text{fm}^{-1}$ in (b), and $V_0 = 6~\text{%
fm}^{-2}$, $r_0 = 4~\text{fm}$ in (c). The mass of particle is adopted as $%
M=0.5~\text{fm}^{-1}$.}
\end{figure}

\begin{figure}[tbp]
\includegraphics[width=12cm]{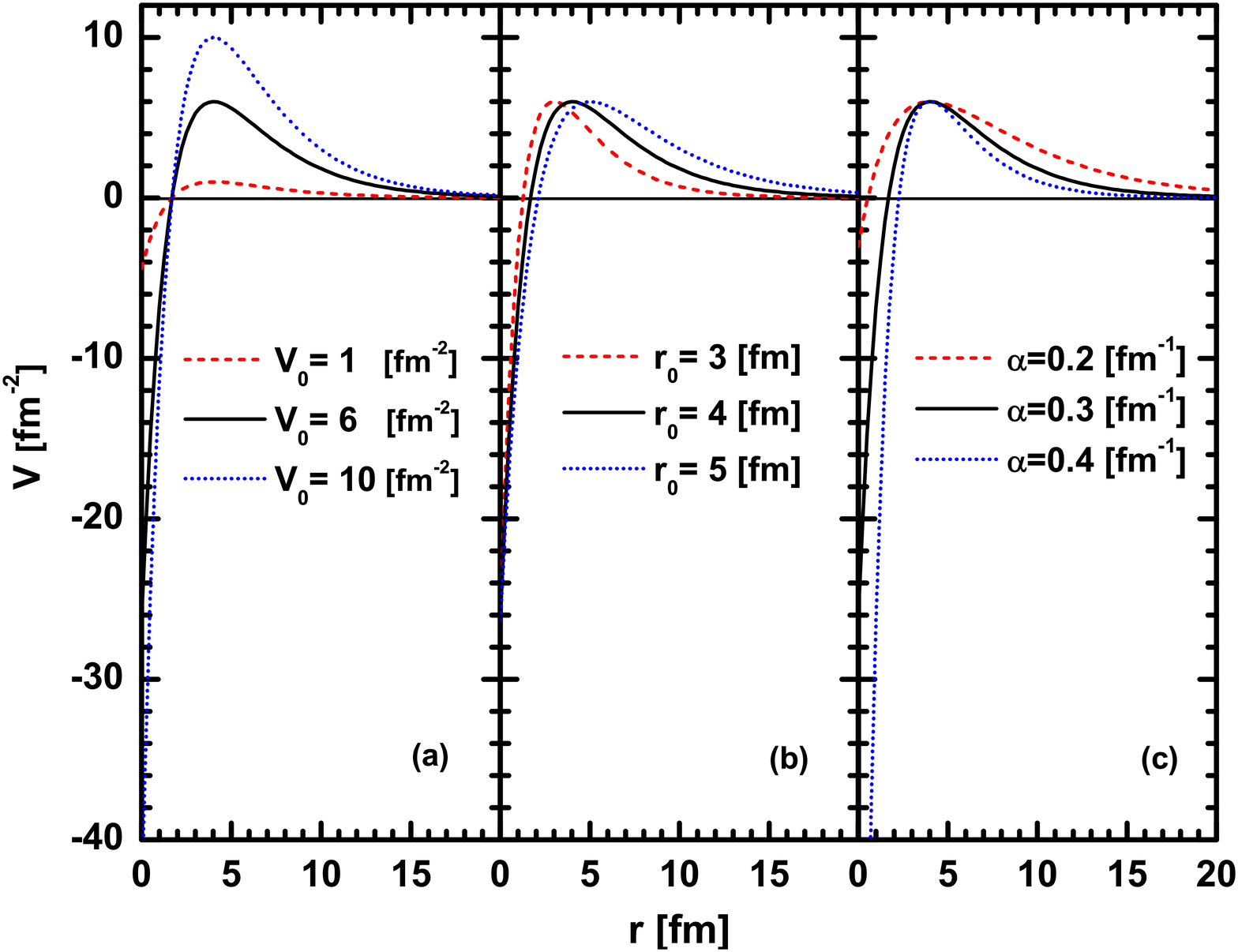}
\caption{(Color online) The Morse potential model for (a) various strength
of the potential $V_0=1, 6, 10~\text{fm}^{-2}$ along with $r_0=4~\text{fm}$
and $\protect\alpha = 0.3~\text{fm}^{-1}$, (b) various equilibrium
intermolecular distance $r_0=3, 4, 5~\text{fm}$ along with $V_0 = 6~\text{fm}%
^{-2}$ and $\protect\alpha = 0.3~\text{fm }^{-1}$, and (c) various positive
number $\protect\alpha = 0.2, 0.3, 0.4~\text{fm}^{-1}$ along with $V_0 = 6~%
\text{fm}^{-2}$ and $r_0 = 4.0~\text{fm}$.}
\end{figure}

\begin{figure}[tbp]
\includegraphics[width=12cm]{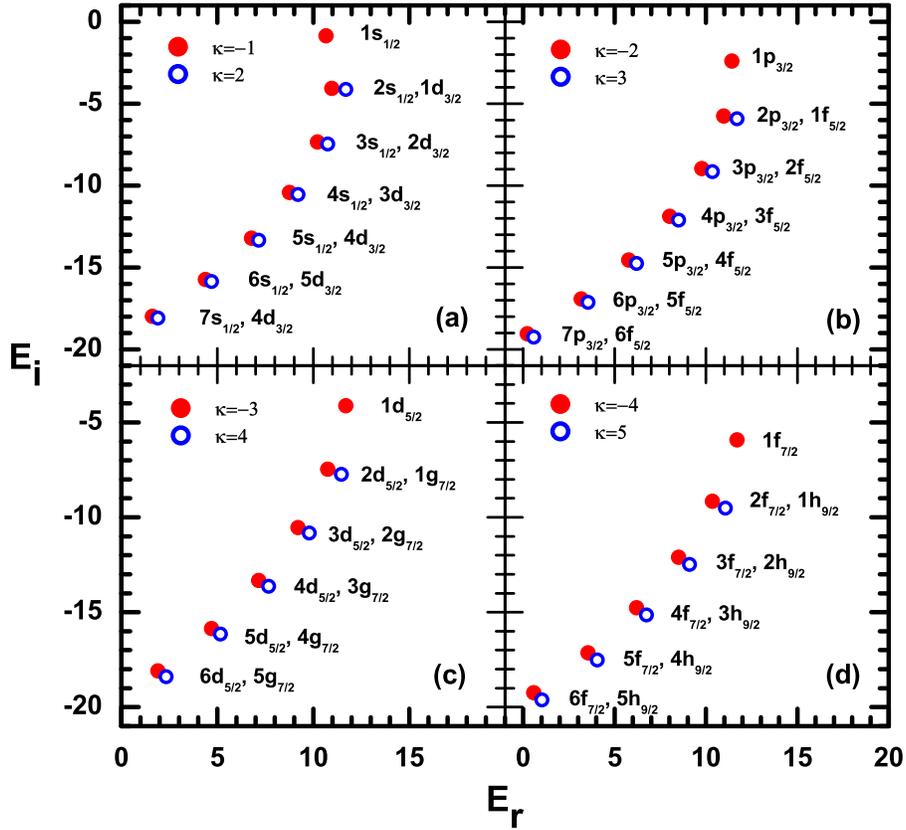}
\caption{(Color online) Pseudospin doublets in the complex energy surface
for the Morse potential with the parameters $V_0 = 10$, $r_0 = 1$, and $%
\protect\alpha = 0.5$ in the atomic units $\hbar=m=1$. }
\end{figure}

\begin{figure}[tbp]
\includegraphics[width=14cm]{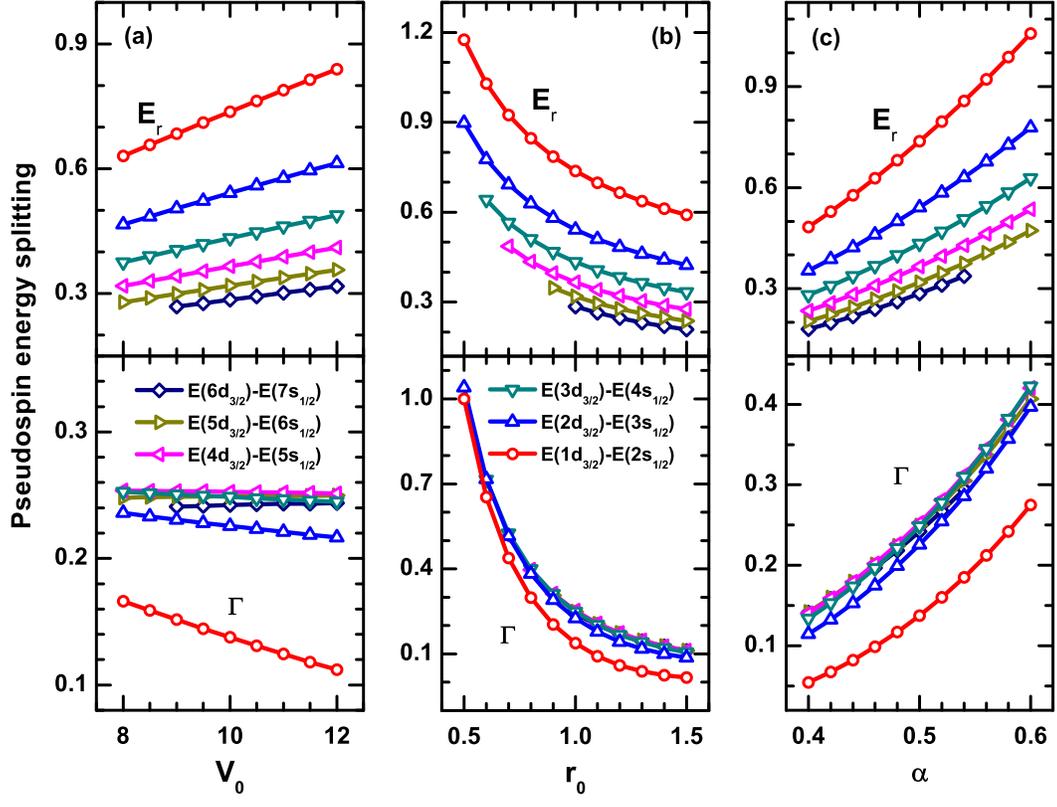}
\caption{(Color online) Pseudospin energy and width splittings as a function
of every potential parameter for the resonant states shown in Fig.~4(a). The
data corresponding to the variables $V_0$, $r_0$, and $\protect\alpha$ are
respectively displayed in the subfigures (a), (b), and (c) with the other
parameters fixed to $r_0 = 1$, $\protect\alpha = 0.5$ in (a), $V_0 = 10$, $%
\protect\alpha = 0.5$ in (b), and $V_0 = 10$, $r_0 = 1$ in (c) in the atomic
units $\hbar=m=1$.}
\end{figure}

\end{document}